\documentclass[11pt,a4paper]{article}

\usepackage{url}
\usepackage{jcappub}
\usepackage{bm}
\usepackage{amsfonts}
\usepackage{subfigure}
\usepackage{graphicx}
\usepackage{upgreek}

\newcommand{\be}{\begin{equation}}
\newcommand{\ee}{\end{equation}}
\newcommand{\bea}{\begin{eqnarray}}
\newcommand{\eea}{\end{eqnarray}}

\newcommand{\beq}{\begin{equation}}
\newcommand{\eeq}{\end{equation}}

\graphicspath{{figure/}}

\newcommand{\ccEps}{Mg(SO$_4)\!\cdot\!7($H$_2$O)}
\newcommand{\ccGyp}{Ca(SO$_4)\!\cdot\!2$(H$_2$O)}
\newcommand{\ccSin}{CaCl$_2 \! \cdot\! 2($H$_2$O)}

\begin{document}

\subheader{\hfill IFIC/22-11}

\title{Rocks, Water and Noble Liquids: Unfolding the Flavor Contents of Supernova Neutrinos} 

\author[a]{Sebastian Baum,}
\author[b,c]{Francesco Capozzi,}
\author[b,d]{Shunsaku Horiuchi}

\emailAdd{sbaum@stanford.edu} \emailAdd{fcapozzi@ific.uv.es} \emailAdd{horiuchi@vt.edu} 

\affiliation[a]{Stanford Institute for Theoretical Physics, Stanford University, Stanford, CA 94305, USA}
\affiliation[b]{Center for Neutrino Physics, Department of Physics, Virginia Tech, Blacksburg, VA 24061, USA}
\affiliation[c]{Instituto de F\'isica Corpuscular, Universidad de Valencia \& CSIC, Edificio Institutos de Investigaci\'on, Calle Catedr\'atico Jos\'e Beltr\'an 2, 46980 Paterna, Spain}
\affiliation[d]{Kavli IPMU (WPI), UTIAS, The University of Tokyo, Kashiwa, Chiba 277-8583, Japan}

\abstract{Measuring core-collapse supernova neutrinos, both from individual supernovae within the Milky Way and from past core collapses throughout the Universe (the diffuse supernova neutrino background, or DSNB), is one of the main goals of current and next generation neutrino experiments. Detecting the heavy-lepton flavor (muon and tau types, collectively $\nu_x$) component of the flux is particularly challenging due to small statistics and large backgrounds. While the next galactic neutrino burst will be observed in a plethora of neutrino channels, allowing to measure a small number of $\nu_x$ events, only upper limits are anticipated for the diffuse $\nu_x$ flux even after decades of data taking with conventional detectors. However, paleo-detectors could measure the time-integrated flux of neutrinos from galactic core-collapse supernovae via flavor-blind neutral current interactions. In this work, we show how combining a measurement of the average galactic core-collapse supernova flux with paleo detectors and measurements of the DSNB electron-type neutrino fluxes with the next-generation water Cherenkov detector Hyper-Kamiokande and the liquid noble gas detector DUNE will allow to determine the mean supernova $\nu_x$ flux parameters with precision of order ten percent.
}

\maketitle

\section{Introduction}
\label{sec:intro}

Supernovae play an important role for the workings of our Universe as well as for our understanding thereof. The rapid injection of energy from supernovae is crucial for the properties of galaxies, and the extreme environment of a supernova makes them the birthplaces of much of the heavy elements in our Universe as well as prolific sources of energetic cosmic rays. Neutrinos play a crucial role in the explosion mechanisms of supernovae, not least because the matter density during the initial stages of a core-collapse supernova is so high that neutrinos are the only (known) particles which can escape this environment. While core-collapse supernovae produce enormous fluxes of all flavors of neutrinos, to date, the only supernova neutrinos measured by terrestrial experiments are the historical $\bar{\nu}_e$ events detected from SN1987A~\cite{Hirata:1987hu, Bionta:1987qt, Aglietta:1987, Alekseev:1988gp}. Detecting all flavors of supernova neutrinos is an important goal, for example to infer the total energetics of the collapse, to test complex neutrino oscillation phenomena, and to search for physics beyond the Standard Model~\cite{Horiuchi:2018ofe}. 

In the upcoming decades, the next generation of neutrino detectors, in particular Hyper-Kamiokande~\cite{Hyper-Kamiokande:2018ofw} and DUNE~\cite{DUNE:2020ypp}, will provide unprecedented sensitivity to supernova neutrinos. Utilizing charged-current interactions, Hyper-Kamiokande is predominantly sensitive to $\bar{\nu}_e$ neutrinos while DUNE is mostly sensitive to the $\nu_e$ fluxes. In order to detect heavy lepton flavor neutrinos $\nu_x = \{ \nu_\mu, \bar{\nu}_\mu, \nu_\tau, \bar{\nu}_\tau\}$, one must rely on processes with significantly smaller event rates. One strategy to detect $\nu_x$ is to use scattering on protons in scintillator detectors~\cite{Beacom:2002hs,Dasgupta:2011wg,Li:2017dbg,Li:2019qxi,Chauhan:2021snf}. Another strategy is to exploit flavor-blind channels, for example, elastic scattering events in large direct dark matter detection experiments, and subtract the (anti-)electron-type neutrino contribution by merging data sets. 

If next-generation neutrino detectors are operational at the time of the next galactic core-collapse event, supernova neutrinos of all flavors will be measured with relatively large statistics - Hyper-Kamiokande, DUNE and JUNO, would detect $\mathcal{O}(10^5)$ $\bar{\nu}_e$, $\mathcal{O}(10^3)$ $\nu_e$ and $\mathcal{O}(10^2)$ $\nu_x$ events, respectively. In addition, the current generation of tonne-scale dark matter detectors such as XENONnT, LZ, or PandaX-4T, would make flavor-blind detections of neutrinos from such a galactic core-collapse event~\cite{Horowitz:2003cz, Lang:2016zhv}. While neutrinos from the next galactic core-collapse supernova will teach us invaluable lessons about neutrinos and supernova physics, this approach also comes with drawbacks: The galactic core-collapse supernova rate is estimated to be $2-3$ per century \cite{Cappellaro:2003eg,Diehl:2006cf,Strumia:2006db,Li:2011,Botticella:2011nd,Adams:2013ana} and the only remedy is patience. Furthermore, one would measure the neutrinos from an individual core-collapse supernova which may well be atypical. Both of these issues are mitigated by attempting to detect neutrinos from the Diffuse Supernova Neutrino Background (DSNB), the neutrino flux from distant core-collapse supernovae throughout the Universe~\cite{Beacom:2010kk,Lunardini:2010ab}. However, detecting the DSNB is challenging due to its relatively small neutrino flux and because the DSNB spectrum is significantly redshifted - the DSNB flux is dominated by supernovae at redshifts $z \sim 1-2$. The next generation of neutrino detectors, in particular Hyper-Kamiokande and DUNE, are expected to detect $\bar{\nu}_e$ and $\nu_e$ neutrinos from the DSNB and measure the corresponding neutrino flux parameters. However, no (conventional) neutral-current detector with the required combination of exposure and energy threshold is available to measure the $\nu_x$ fluxes. Even future direct dark matter detectors such as DARWIN~\cite{DARWIN:2016hyl} are expected to set only upper limits on the DSNB $\nu_x$ flux roughly an order of magnitude above the expected flux~\cite{Suliga:2021hek}.

In this work, we show the prospects of {\it paleo detectors} to measure the $\nu_x$ supernova neutrino flux. Rather than operating a real-time laboratory experiment to search for neutrinos, paleo detectors~\cite{Baum:2018tfw,Drukier:2018pdy,Baum:2021jak} would use nuclear damage tracks recorded in natural minerals of geological ages (100\,Myr$-$1\,Gyr) to search for neutrino-induced nuclear recoils~\cite{Baum:2019fqm}. Many minerals are excellent solid state track detectors~\cite{Fleischer:1964,Fleischer383,Fleischer:1965yv,GUO2012233} and retain the damage tracks caused by nuclear recoils for timescales which can exceed the age of the Earth by many orders of magnitude. Modern microscopy techniques such as Small Angle X-Ray scattering~\cite{RODRIGUEZ2014150,SAXS3d,SAXSres} potentially allow for the readout of these damage tracks in macroscopic samples. Paleo detectors have been considered to study atmospheric \cite{Jordan:2020gxx} and solar neutrinos \cite{Collar:1995aw, Tapia-Arellano:2021cml} in addition to supernova neutrinos \cite{Baum:2019fqm}.

There are two crucial differences between detecting supernova neutrinos in paleo detectors and in Hyper-Kaomiokande/DUNE: First, the geological time scales paleo detectors record neutrino-interactions over are many orders of magnitude longer than the (inverse) supernova rate of the Milky Way ($2-3$ per century~\cite{Cappellaro:2003eg,Diehl:2006cf,Strumia:2006db,Li:2011,Botticella:2011nd,Adams:2013ana}). Physically then, paleo detectors are sensitive to the mean supernova neutrino properties from many past core collapses, similar to the DSNB modulo different distances. When averaged over timescales much longer than a century, the supernova neutrino flux on Earth is no longer dominated by the DSNB, but by neutrinos from supernovae in our galaxy, and this time-averaged flux is a factor of order 100 larger than the DSNB flux~\cite{Baum:2019fqm}. Second, paleo detectors would detect supernova neutrinos via Coherent Elastic neutrino-Nucleus Scattering (CE$\nu$NS), a process mediated by neutral currents and hence sensitive to the total neutrino flux rather than the $\bar\nu_e$/$\nu_e$ fluxes Hyper-Kamiokande/DUNE would measure. In this regard, paleo detectors utilize the same interaction physics as direct dark matter experiments. Combined with the significantly larger time-integrated galactic supernova neutrino flux, paleo detectors offer a unique opportunity to measure all flavors of neutrinos emitted from a population of past core-collapse events. 

In order to extract the $\nu_x$ flux parameters, a measurement of the total galactic supernova neutrino flux in paleo detectors must be combined with flavor-sensitive measurements of supernova neutrinos. To this end, we consider a combination of the paleo-detector measurement with measurements of the DSNB $\bar\nu_e$/$\nu_e$ parameters with Hyper-Kamiokande/DUNE. The main background for supernova neutrinos in paleo detectors stems from radiogenic neutrons originating from the $^{238}$U decay chain, and we project the precision to which the $\nu_x$ flux parameters can be extracted for different assumptions on the uranium concentration in paleo-detector samples. We find that for a $^{238}$U concentration of $10^{-11}\,$g/g ($10^{-12}\,$g/g), the average energy as well as the total energy of the $\nu_x$ flux can be measured with few tens of percent ($\lesssim 10\%$) precision.

The remainder of this work is structured as follows: In section~\ref{sec:Paleo}, we discuss the calculation of the time-averaged flux of neutrinos from galactic supernovae at Earth, the calculation of the DSNB and the calculation of the neutrino-induced signal in paleo detectors. A discussion of the most relevant background sources in paleo detectors can be found in section~\ref{sec:Paleo-bkg}. In section~\ref{sec:Analysis} we explain the statistical analysis we use to extract the neutrino parameters from mock data in Hyper-Kamiokande, DUNE, and paleo detectors. We show our results for the projected precision of the reconstruction of these parameters in section~\ref{sec:Results}. In section~\ref{sec:Conclusions}, we summarize and discuss our findings.

\section{Paleo detectors as (supernova) neutrino detectors}
\label{sec:Paleo}

In this section, we discuss the signal from supernova neutrinos in paleo detectors as well as the most relevant backgrounds. For a more detailed discussion of paleo detectors as (CE$\nu$NS) neutrino detectors, see Ref.~\cite{Baum:2019fqm}, and for discussions of the relevant background sources in paleo detectors see also Refs.~\cite{Drukier:2018pdy,Baum:2021jak}. 

We model the spectrum of the neutrino species $\nu_i$ from a supernova as a pinched Fermi-Dirac distribution~\cite{Keil:2002in}
\begin{equation}\label{eq:nu_spec}
   \left(\frac{d n}{d E}\right)_{\nu_i} = E_\nu^{\rm tot} \frac{\left(1+\alpha\right)^{1+\alpha}}{\Gamma(1+\alpha)} \frac{E^{\alpha}}{\langle E_\nu\rangle^{2+\alpha}} e^{\left[ -\left(1+\alpha\right) \frac{E}{\langle E_\nu \rangle} \right]} \;,
\end{equation}
where $E_\nu^{\rm tot}$ is the total energy radiated in $\nu_i$, $\langle E_\nu \rangle$ is the average energy of the neutrinos, and $\alpha$ is a (dimensionless) spectral shape parameter. For flavor-blind interactions such as CE$\nu$NS, the relevant neutrino flux is the sum over all flavors,
\begin{equation}\label{eq:flavor_sum}
  \frac{d n}{d E_\nu} = \left(\frac{d n}{d E}\right)_{\nu_e} + \left(\frac{d n}{d E}\right)_{\bar\nu_e} + 4 \left(\frac{d n}{d E}\right)_{\nu_x} \;.
\end{equation}
The time-averaged neutrino spectrum from core-collapse supernovae in our Galaxy is then obtained by integrating over the probability density $f(R_E)$ for a supernova to occur at a distance $R_E$ from Earth and by multiplying with the galactic core-collapse supernova rate, $\dot{N}_{\rm CC}^{\rm gal}$,
\begin{equation}
  \left( \frac{d \phi}{d E_\nu} \right)^{\rm gal} = \dot{N}_{\rm CC}^{\rm gal} \frac{d n}{d E_\nu} \int_0^\infty d R_E \, \frac{f(R_E)}{4 \pi R_E^2} \;.
\end{equation}
For $f(R_E)$ we assume that supernovae occur predominantly in the stellar disk and model the distribution of core-collapse supernovae as a double-exponential in galactocentric coordinates,
\begin{equation}
  \rho \propto e^{- R/R_d} e^{-\left|z\right|/H} \;,
\end{equation}
where $R$ is the galactocentric radius, $z$ the height above the galactic plane, and we use $R_d = 2.9\,$kpc and $H = 95\,$pc~\cite{Adams:2013ana}. 

Similarly, the DSNB neutrino flux is obtained by integrating over the volumetric cosmic supernova rate, $\dot{n}_{\rm CC}^{\rm cosmo}$,~\cite{Beacom:2010kk,Lunardini:2010ab}
\begin{equation} \label{eq:DSNB}
    \left( \frac{d \phi}{d E_\nu} \right)^{\rm DSNB} = \int_0^\infty \frac{dz}{H_0 \sqrt{\Omega_\Lambda + \Omega_m \left(1+z\right)^3} } \dot{n}_{\rm CC}^{\rm cosmo}(z) \frac{d n}{d E'_\nu} \;,
\end{equation}
where $H_0 \simeq 67\,$km/s/Mpc is the Hubble constant, $\Omega_\Lambda \simeq 0.68$ and $\Omega_m \simeq 0.32$ are the cosmological density parameters for dark energy and matter, respectively, and $z$ is the cosmological redshift. Note that the supernova neutrino spectrum, $d n/d E_\nu$, must be evaluated at the redshifted neutrino energy $E_\nu' = (1+z)E_\nu$. The cosmic supernova rate is related to the (volumetric) star formation rate $\dot{\rho}_*(z)$ via the conversion factor
\begin{equation}\label{eq:rate}
    \dot{n}_{\rm CC}^{\rm cosmo}(z) = \dot{\rho}_*(z)\frac{\int_{M_{\rm min} = 8}^{100}\psi(M)\,dM}{\int_{0.1}^{100} M \psi(M) \,dM},
\end{equation}
which is equivalent to assuming that all massive stars with mass above $M_{\rm min} = 8 M_\odot$ undergo core collapse. This is a reasonable assumption for our estimates (see, e.g., discussions in \cite{Horiuchi:2008jz}) -- although a fraction of massive stars are expected to collapse to black holes, with systematically higher temperature neutrino emissions \cite{Horiuchi:2017qja}, we neglect their contribution in this work. Firstly, their occurrence rate remains highly uncertain, and secondly, their contributions would enhance the detectability of the DSNB so we remain conservative by omitting their occurrence. The star-formation rate is observationally measured~\cite{Hopkins:2006bw,Madau:2014bja}, especially well in the low redshift range of importance for estimating the DSNB, and we adopt the functional fit described in, e.g., Refs.~\cite{Horiuchi:2017qja,Horiuchi:2020jnc}. Note that alternatively one could also take $\dot{n}_{\rm CC}^{\rm cosmo}(z)$ directly from supernova surveys, see for example Ref.~\cite{Strolger:2015kra}. However, the limited depth of current supernova surveys means that in the most relevant redshift range, $ z \lesssim 2$, the errors of such direct measurements of $\dot{n}_{\rm CC}^{\rm cosmo}(z)$ are larger than those of inferring $\dot{n}_{\rm CC}^{\rm cosmo}(z)$ from measurements of the cosmic star formation rate via eq.~\eqref{eq:rate}. 

Supernova neutrinos (predominantly\footnote{Additional signals arise via quasi-elastic charged-current interactions and, for more energetic neutrinos, via (deep) inelastic neutrino-nucleus interactions~\cite{Jordan:2020gxx}. However, the contribution from quasi-elastic interactions is suppressed compared to the CE$\nu$NS one by the lack of coherent enhancement, and the contribution of inelastic interactions of more enegetic neutrinos is suppressed by the quickly falling neutrino flux.}) induce signals in paleo detectors via CE$\nu$NS interactions. The differential recoil spectrum (per unit target mass) for a target nucleus $T$ to recoil with energy $E_R$ is given by~\cite{Billard:2013qya,OHare:2016pjy}
\begin{equation}
  \left( \frac{d R}{d E_R}\right)_T = \frac{1}{m_T} \int_{E_\nu^{\rm min}} d E_\nu\, \frac{d \sigma}{d E_R} \frac{d \phi}{d E_\nu} \;,
\end{equation}
where $m_T$ is the mass of $T$ and $E_\nu^{\rm min} = \sqrt{m_T E_R/2}$ is the minimum neutrino required to induce a nuclear recoil with $E_R$. The differential CE$\nu$NS cross section is
\begin{equation}
  \frac{d \sigma}{d E_R}(E_R, E_\nu) = \frac{G_F^2}{4\pi} Q_W^2 m_T \left( 1 - \frac{m_T E_R}{2 E_\nu^2} \right) F^2(E_R) \;,
\end{equation}
with the Fermi coupling constant $G_F$, the nuclear form factor $F(E_R)$\footnote{We use the Helm form factor~\cite{Helm:1956zz} as in Ref.~\cite{Baum:2019fqm}}, and
\begin{equation}
  Q_W \equiv \left( A_T - Z_T \right) - \left( 1 - 4 \sin^2 \theta_W \right) Z_T \;,
\end{equation}
where $\theta_W$ is the weak mixing angle, and $A_T$ ($Z_T$) the number of nucleons (protons) in $T$.

The observable in a paleo detector is the track length spectrum, which is obtained from the recoil energy spectrum by summing over all target nuclei and weighting with the stopping power of $T$ in the material, $d E_R / d x_T$,
\begin{equation}
  \frac{d R}{d x} = \sum_T \xi_T \frac{d E_R}{d x_T} \left( \frac{d R}{d E_R} \right)_T \;,
\end{equation}
where $\xi_T$ is the mass fraction of $T$ in the target material. Similarly, the length $x$ of a track from a nucleus $T$ recoiling with energy $E_R$ is obtained by integrating the stopping power,
\begin{equation}
  x(E_R) = \int_0^{E_R} d E\, \left| \frac{d E}{d x_T} \right|^{-1} \;.
\end{equation} 
We use \texttt{SRIM}~\cite{Ziegler:1985,Ziegler:2010} to calculate stopping powers in the target materials used in this work, epsomite [\ccEps], gypsum [\ccGyp], and sinjarite [\ccSin].

\subsection{Background Sources} \label{sec:Paleo-bkg}
Let us briefly discuss the most relevant background sources for supernova neutrino searches in paleo detectors. These backgrounds have been discussed in detail in Refs.~\cite{Drukier:2018pdy,Baum:2019fqm} see also Ref.~\cite{Baum:2021jak}. Note that all relevant backgrounds stem from nuclear recoils. Natural defects in minerals are either single-site or span across the entire mono-crystalline volume, and thus do not resemble the damage tracks induced by CE$\nu$NS of supernova neutrinos. 
\begin{itemize}
    \item {\it Cosmogenics:} In order to suppress cosmic-ray induced backgrounds, minerals used as paleo detectors must have been shielded by a sufficiently large overburden for the entire time they have been recording nuclear damage tracks. The modest amounts of target materials required for paleo detectors, at most a few kg, can be sourced from much greater depths than those of existing underground laboratories. For example, boreholes drilled for geological R\&D or oil exploration are promising sources of samples. For an overburden of $\gtrsim 5\,$km rock, cosmogenic backgrounds, in particular the cosmogenic muon induced neutron flux, are suppressed to a negligible level. 
    
    \item {\it Radiogenics:} Any mineral used as a paleo detector will be contaminated by trace amounts of radioactive material. In order to mitigate the associated radiogenic backgrounds, it is crucial to use as radiopure minerals as possible as paleo detectors. The most important radioactive isotope in natural minerals is $^{238}$U. Minerals formed in the Earth's crust have typical $^{238}$U concentrations of $C^{238} \sim 10^{-6}\,$g/g, which would lead to prohibitively large radiogenic backgrounds. The $^{238}$U concentration in so-called Ultra Basic Rocks (UBRs), formed from the material of the Earth's mantle, and in Marine Evaporites (MEs), salts formed from sea water, are much lower, making them attractive as paleo detectors, see Refs.~\cite{Thomson:1954,Condie:1957,Adams:1959,Seitz:1974,Dean:1978,Yui:1998,Sanford:2013} as well as the discussion in Refs.~\cite{Drukier:2018pdy,Baum:2019fqm}. As in previous work on paleo detectors~\cite{Baum:2018tfw,Drukier:2018pdy,Edwards:2018hcf,Baum:2019fqm,Jordan:2020gxx,Baum:2021jak}, we will assume benchmark $^{238}$U concentrations of $C^{238} = 10^{-11}\,$g/g for the ME examples epsomite [\ccEps], gypsum [\ccGyp], and sinjarite [\ccSin] we use as target materials in this work. We will also show prospects for measuring supernova neutrinos for a more optimistic assumption of $C^{238} = 10^{-12}\,$g/g. The most relevant radiogenic background in paleo detectors are nuclear recoils induced by the elastic scattering of radiogenic neutrons off the nuclei constituting the mineral. These neutrons are produced by spontaneous fission of heavy radiogenic isotopes as well as by $(\alpha,n)$-reactions of the $\alpha$-particles from the $^{238}$U decay chain. We use \texttt{SOURCES-4A}~\cite{sources4a:1999} to model these neutron spectra, and calculate the induced nuclear recoil spectrum using \texttt{TENDL-2017}~\cite{Koning:2012zqy,Rochman:2016,Sublet:2015,Fleming:2015} neutron-nucleus cross sections as tabulated in the \texttt{JANIS4.0} database~\cite{Soppera:2014zsj}.\footnote{We take only elastic neutron-nucleus scattering into account; this yields a conservative estimate of the background because neutrons typically lose a larger fraction of their energies through inelastic processes than in elastic scattering. Note also that our Monte Carlo simulation of the nuclear recoils induced by radiogenic neutrons has been validated with a calculation of the nuclear recoils induced by the same neutron spectra with \texttt{FLUKA}~\cite{Ferrari:2005zk,Bohlen:2014buj,NUNDIS} in Ref.~\cite{Jordan:2020gxx} for the particular case of a halite target.}
    
    \item {\it Astrophysical neutrinos:} Just like supernova neutrinos, neutrinos from other sources give rise to nuclear recoils and, in turn, nuclear damage tracks in paleo detectors. The most relevant backgrounds to supernova neutrino induced recoils are solar and atmospheric neutrinos. We take the corresponding neutrino fluxes from Ref.~\cite{OHare:2020lva} and model the corresponding nuclear recoil spectrum as for the supernova neutrinos described above.
\end{itemize}

\begin{figure}
    \centering
    \includegraphics[width=0.7\linewidth] {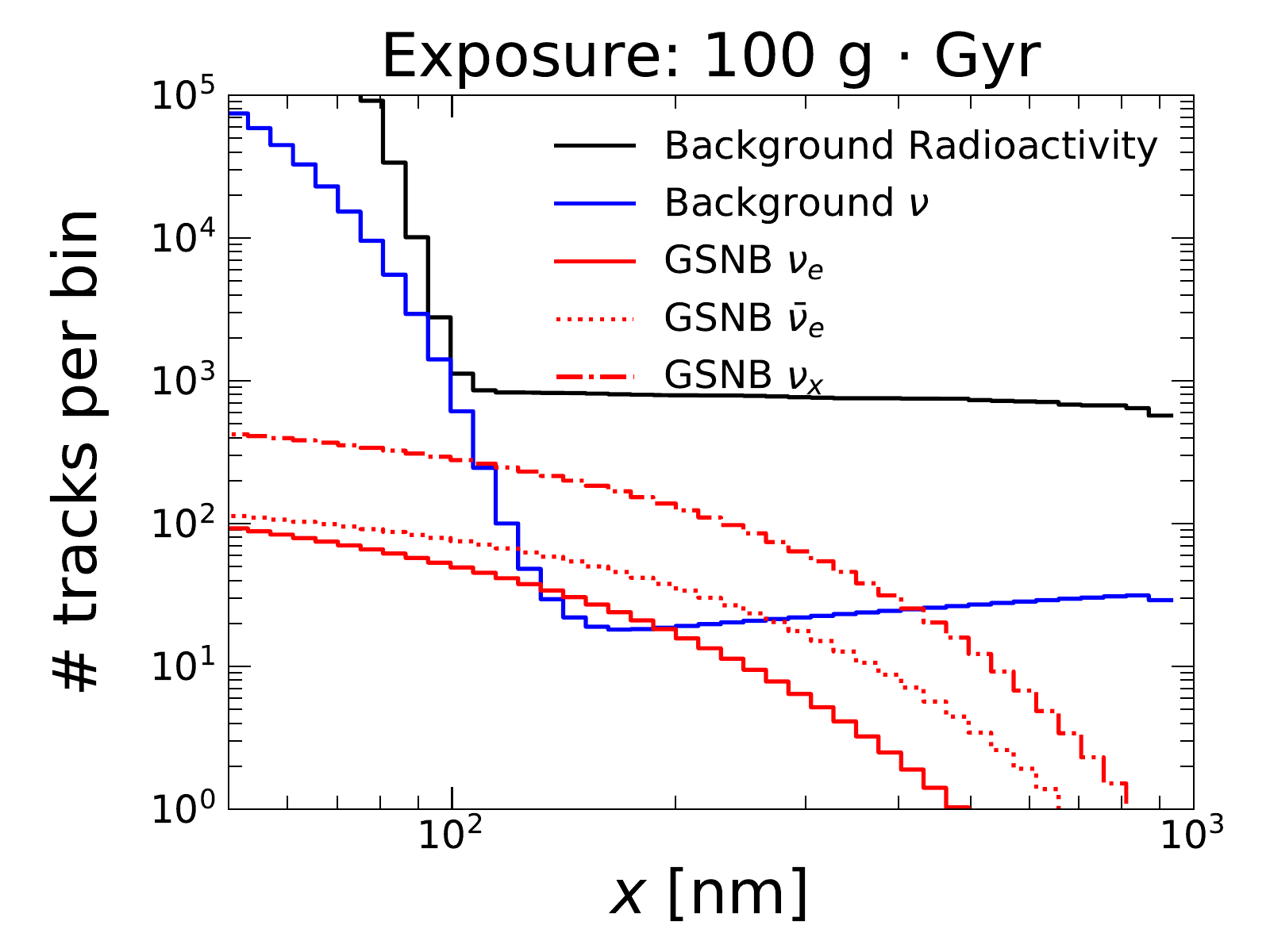}
    \caption{Number of events per bins of track length in a Epsomite [\ccEps] paleo detector with 100\,gram\,Gyr exposure. Throughout this work, we use 70 log-spaced bins in the range $7.5\,{\rm nm} \leq x \leq 10^3\,$nm. The red lines show the events induced by CE$\nu$NS of neutrinos from core-collapse supernovae in our galaxy. The blue line indicates the spectrum induced by scattering of solar, atmospheric, and DSNB neutrinos, and the black line shows the spectrum induced by radiogenic neutrons. As we show in this work, by utilizing the entire spectrum the galactic supernova signal and its properties can be measured although it produces fewer tracks than the backgrounds for all track lengths.}
    \label{fig:spectra}
\end{figure}

In figure~\ref{fig:spectra} we show the signal induced by the coherent scattering of neutrinos from supernovae in our Galaxy in epsomite [\ccEps] together with the backgrounds induced by other astrophysical neutrinos and radiogenic neutrons, assuming $C^{238}=10^{-11}\,$g/g. Note that we have included contributions from solar and atmospheric neutrinos as well as from the DSNB in the spectrum labeled ``$\nu$ bkg''. As we will see, the properties of the galactic supernova neutrino flux can be inferred in a paleo detector despite the fact that the overall normalization of the associated track length spectrum is smaller than that of the backgrounds. This is possible thanks to the enormous exposure achievable in paleo detectors: Reading out 100\,g of a mineral which has been recording damage tracks for 1\,Gyr corresponds to an exposure of $\varepsilon = 100\,{\rm g\,Gyr} = 100\,$kt\,yr. For such an exposure, there would be about $2.8\times 10^3$ $\nu_x$ tracks with length between 100 and 500\,nm. In the same range, there would be $1.8\times 10^5$ tracks from background events for our fiducial assumption of $C^{238} = 10^{-11}\,$g/g. Therefore, considering only statistical uncertainty, the $\nu_x$ signal can be observed with a statistical significance of $2.8\times 10^3/\sqrt{1.8\times 10^5}=6.6\sigma$. For the same exposure in gypsum [\ccGyp] and sinjarite [\ccSin], there would be a significance of $3\sigma$ and $2\sigma$, respectively. Note that even for $^{238}$U concentrations as large as $C^{238}=5\times 10^{-11}\,$g/g, the (statistical) significance for seeing $\nu_x$ induced tracks in an epsomite paleo detector would still be $3\sigma$. 

\section{Analysis}
\label{sec:Analysis}

In order to calculate the track length distributions in paleo detectors for both supernova neutrino and background events we use the public code presented in Ref.~\cite{Baum:2021jak}. In our analysis we employ three rock samples: Epsomite [\ccEps], gypsum [\ccGyp] and sinjarite [\ccSin]; these minerals provide particularly good detection statistics for supernova neutrinos. We assume that 100\,g of each material can be read out with track length resolution of 15\,nm.\footnote{In order to include the effects of finite track length resolution, we compute the number of tracks in the $i$-th bin with reconstructed length $x \in \left[ x_i^{\rm min}, x_i^{\rm max} \right]$ in a sample of mass $M$ which has been recording tracks for a time $t_{\rm age}$ as $N_i = M \times t_{\rm age} \int dx'\,W(x';x_i^{\rm min},x_i^{\rm max}) \left( dR/dx' \right)$, where $dR/dx'$ is the true track length spectrum and $W$ is a window function, see, for example, Ref.~\cite{Baum:2021jak}.} Such a sample size and readout resolution may, for example, be achievable with small angle X-ray scattering microscopy~\cite{RODRIGUEZ2014150,SAXS3d,SAXSres}, see the discussion in Ref.~\cite{Drukier:2018pdy}. We consider both a default $^{238}$U concentration of $C^{238}=10^{-11}\,$g/g and a more optimistic one of $C^{238}=10^{-12}\,$g/g. These choices are summarized in table~\ref{tab:1}. Concerning the neutrino energy spectrum, the parameters used for generating our set of mock data are given in table~\ref{tab:2}. These have been extracted from figure~1 in \cite{Bollig:2020phc}, where the first 3D supernova simulation up to 7 seconds post bounce has been presented, providing a more reliable estimate of the total energies of each neutrino species.  In particular, the values for the average energies reported in table~\ref{tab:2} refer to the time-averaged quantities reported in figure~1 of \cite{Bollig:2020phc}, whereas the total energies are obtained by integrating over time. 

\begin{table}
\centering
\caption{Summary of paleo detector properties used in our simulation.}
\begin{tabular}{| c | c |}
\hline\hline
 Track Resolution & $15\,$nm \\ \hline
    Track Length Min & $7.5\,$nm \\ \hline
    Track Length Max & $1000\,$nm \\ \hline
    Number of Log Bins & 70 \\ \hline
    Rock Sample Mass & $100\,$g \\ \hline
    Rock Sample Age & $1\,$Gyr \\ \hline 
    Default Uranium Concentration ($C^{238}$) & $10^{-11}\,$g/g \\ \hline
    Optimistic Uranium Concentration ($C^{238}$) & $10^{-12}\,$g/g \\ \hline\hline
\end{tabular}
\label{tab:1}
\end{table}

\begin{table}
\centering
\caption{Summary of neutrino properties used to generate our set of mock data. Note that $E^{\rm tot}_{\nu_x}$ refers to the total energy of one of the four $\nu_x$ neutrino species ($\nu_\mu$, $\bar{\nu}_\mu$, $\nu_\tau$, $\bar{\nu}_\tau$), see eq.~\eqref{eq:flavor_sum}, and $\alpha_\nu$ is the pinching-parameter of the energy distribution of neutrinos~\cite{Keil:2002in}, see eq.~\eqref{eq:nu_spec}.}
\begin{tabular}{| c | c |}
\hline\hline
 $\langle E_{\nu_e}\rangle$ & $11.2$ MeV \\ \hline
    $\langle E_{\bar{\nu}_e}\rangle$ & $13.5$ MeV \\ \hline
     $\langle E_{\nu_x}\rangle$ & $13.4$ MeV\\ \hline
    $E^{\rm tot}_{\nu_e}$ & $6.8\times10^{52}$ erg \\ \hline
     $E^{\rm tot}_{\bar{\nu}_e}$ & $6.6\times 10^{52}$ erg \\ \hline
    $ E^{\rm tot}_{\nu_x}$ & $6.2\times 10^{52}$ erg \\ \hline
    $\alpha_{\nu_e}=\alpha_{\nu_x}=\alpha_{\bar{\nu}_e}$ & 3 \\\hline
    Galactic Supernova Rate & $0.023$ per year \\ \hline\hline
\end{tabular}
\label{tab:2}
\end{table}

Since the contribution of galactic supernova neutrinos to the track length distribution is sub-dominant with respect to the radiogenic and solar neutrino background, paleo detectors alone cannot (completely) break the degeneracy between the three supernova neutrino components ($\nu_e,\bar{\nu}_e,\nu_x$). To alleviate this problem we include an independent measurement of the DSNB with the future neutrino detectors DUNE~\cite{DUNE:2020lwj,DUNE:2020ypp,DUNE:2020mra,DUNE:2020txw} and Hyper-Kamiokande~\cite{Hyper-Kamiokande:2018ofw}, which will be sensitive to $\nu_e$ and $\bar{\nu}_e$, respectively. Let us note that observations of the neutrinos from a single future galactic supernova in neutrino and dark matter detectors can, in general, provide much higher precision measurements of the supernova neutrino spectrum than the combined measurement of the DSNB in DUNE and Hyper-Kamiokande and galactic supernova neutrinos in paleo detectors we consider here. However, our approach has two important advantages. First, we do not need to wait for the next galactic core-collapse supernova. Second, our approach would measure neutrinos from a large population of supernovae rather than from one specific (possibly atypical) supernova.  

In order to project the precision with which DUNE and Hyper-Kamiokande can constrain the parameters controlling the supernova neutrino spectrum, we compute the DSNB flux as in eq.~\eqref{eq:DSNB}, except that we use the spectrum from the appropriate single neutrino flavor $\nu_i$ rather than the total neutrino flux, see eqs.~\eqref{eq:nu_spec} and~\eqref{eq:flavor_sum}. For detection at DUNE, we consider a 40\,kton liquid argon detector and detection of electron neutrinos through the charged-current interaction $\nu_e + {\rm Ar} \to e^- + {\rm K}^+$. We adopt the cross section based on the random
phase approximation scheme of Ref.~\cite{Kolbe:2003ys}. Since DUNE is under construction, we assume a detection efficiency of 86\% based on the DUNE design report~\cite{DUNE:2020ypp,Tabrizi:2020vmo}. For detection at Hyper-Kamiokande, we consider a gadolinium-enhanced water detector with fiducial volume of 187\,kton and detection of anti-electron neutrinos through inverse-beta interaction on water $\bar{\nu}_2 + p \to n + e^+$. The cross section is well-known~\cite{Vogel:1999zy,Strumia:2003zx}, and we assume a detection efficiency of 67\%~\cite{Tabrizi:2020vmo}.  We consider a 20-year run time for both Hyper-Kamiokande and DUNE. We note that although other detectors, e.g., JUNO~\cite{JUNO:2015sjr}, have sensitivity to DSNB neutrinos~\cite{Moller:2018kpn}, DUNE and Hyper-Kamiokande will dominate the sensitivity to $\nu_e$ and $\bar{\nu}_e$, respectively. 

In our analysis we perform a scan over a 12-dimensional parameter space. Six parameters correspond to the average and total energies of $\nu_e,\bar{\nu}_e$ and $\nu_x$ (the pinching parameters $\alpha_{\nu_i}$ are fixed to the values in table~\ref{tab:2}). The floating parameters related to paleo detectors are the normalization of the radiogenic and solar neutrino background. Those related to DSNB measurements are the normalization of the backgrounds in DUNE and Hyper-Kamiokande, for which we adopt the spectra modeled in Refs.~\cite{Moller:2018kpn,Tabrizi:2020vmo}. The last two parameters are the galactic supernova rate and the overall normalization of the DSNB signal, which parametrize the uncertainty on the star formation rate. We introduce Gaussian priors with (relative) width of 10\% on the galactic supernova rate and the normalization of the DSNB, and priors with 20\% width on the background normalization in DUNE and Hyper-Kamiokande.  To put our choices of priors on the galactic supernova rate and the DSNB in contex, we note that the uncertainty on the cosmic star formation rate is already at the level of $\sim 20\%$ at the low redshifts most relevant to the DSNB and local core collapses~\cite{Hopkins:2006bw,Beacom:2010kk,Madau:2014bja}. We expect considerable improvements on this uncertainty in the next decade, not least from the upcoming surveys with the James Webb Space Telescope~\cite{Gardner:2006ky} and the Vera C. Rubin Observatory~\cite{LSST:2009}. While the time evolution of the galactic star formation rate on gigayear timescales remains uncertain with tens-of-percent fluctuations (see, e.g., Refs.~\cite{Snaith:2015,Haywood:2016,Mor:2019}), we emphasize that the relevant quantity for our analysis is the local supernova rate integrated over the past $\sim 1\,$Gy rather than the instantaneous rate. Note that we are not including any priors on the normalization of the solar neutrino and radiogenic backgrounds in paleo detectors nor on the total and mean energies of the supernova neutrino fluxes. We scan over the parameter space using a Markov Chain Monte Carlo through the public software \texttt{PyMultinest}~\cite{Feroz:2008xx,Buchner:2014nha}. We define the following likelihood:
\begin{equation}
    \begin{split}
        \log(\mathcal{L})=&-\sum_{i=1}^{N_{\rm bins}^{\rm paleo}}\left[N_i^{\rm th}-N_{i}^{\rm data}+N_{i}^{\rm data}\,\left(\log(N_{i}^{\rm data})-\log(N_{i}^{\rm th}\right)\right]-\frac{1}{2}\sum_{p}\left(\frac{f_p}{\sigma_p}\right)^2\\
        &-\sum_{\rm exp}\sum_{i=1}^{N_{\rm bins}^{\rm exp}}\left[M_{{\rm exp},i}^{\rm th}-M_{{\rm exp},i}^{\rm data}+M_{{\rm exp},i}^{\rm data}\,\left(\log(M_{{\rm exp},i}^{\rm data})-\log(M_{{\rm exp},i}^{\rm th}\right)\right]\,,
    \end{split}
    \label{eq:likelihood}
\end{equation}
where $N_i^{\rm data}$ is the number of simulated data events in the $i$-th track length bin for paleo detectors and $N_i^{\rm th}$ the number of expected events, defined as $N_i^{\rm th}=N_i^{\rm 0,th}(1+\sum_pf_p)$, where the $f_p$ are the free normalization parameters discussed above and $N_i^{\rm 0,th}$ is the expected number of events calculated with the default values of parameters. $M_{{\rm exp},i}^{\rm data,th}$ is defined analogously to  $N_i^{\rm th}$, but refers to DSNB events for exp$=$(Hyper-Kamiokande,DUNE). Concerning the binning, we use 70 log-spaced bins between $7.5\,{\rm nm} $ and $10^3\,$nm.

For the DSNB we take into account also possible neutrino oscillations inside the supernova.\footnote{Since paleo detectors measure supernova neutrinos (predominantly) via CE$\nu$NS, flavor oscillations have no effect on the signal in paleo detectors.} In particular, focusing only on the MSW effect and neglecting collective and turbulence effects, we consider three cases: no oscillations, normal and inverted mass ordering. For normal mass ordering the survival probability of $\nu_e$ is $P(\nu_e\to\nu_e)=0$, whereas the one for $\bar{\nu}_e$ is $P(\bar{\nu}_e\to\bar{\nu}_e)=0.7$. For inverted mass ordering the survival probability of $\nu_e$ is $P(\nu_e\to\nu_e)=0.3$, whereas the one for $\bar{\nu}_e$ is $P(\bar{\nu}_e\to\bar{\nu}_e)=0$. Note that when we report our projected constraints on the neutrino parameters below, neutrino flavors always refer to the those emitted in the supernova explosion, i.e., prior to any flavor oscillations. 

\section{Results}
\label{sec:Results}

\begin{figure*}
\centering
    \text{\scriptsize Only DSNB (HyperK + DUNE), no oscillations}\par
    \includegraphics[trim={0cm 0cm 0cm 6mm},clip,width=0.48\linewidth]{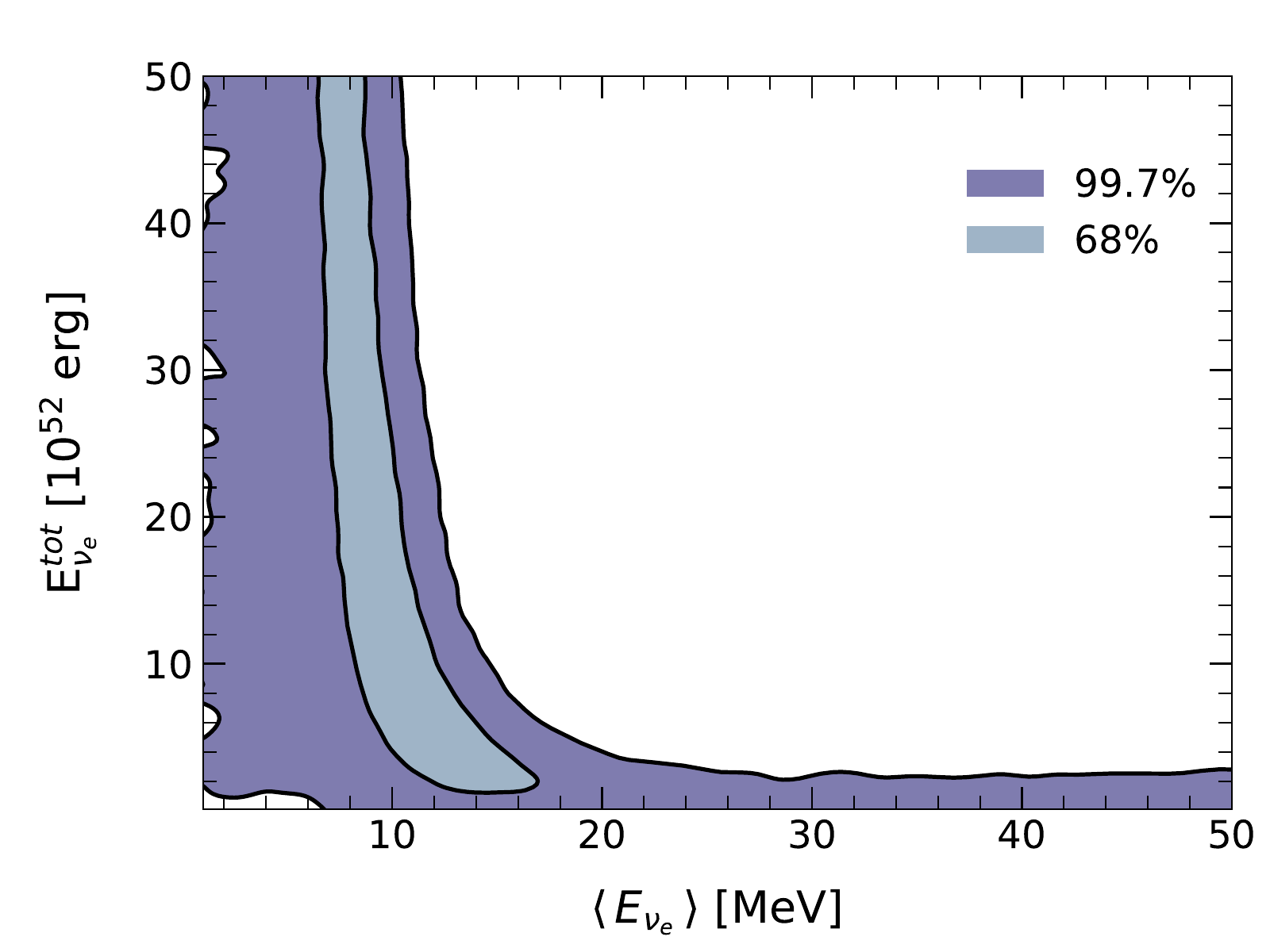}
    \includegraphics[trim={0cm 0cm 0cm 6mm},clip,width=0.48\textwidth]{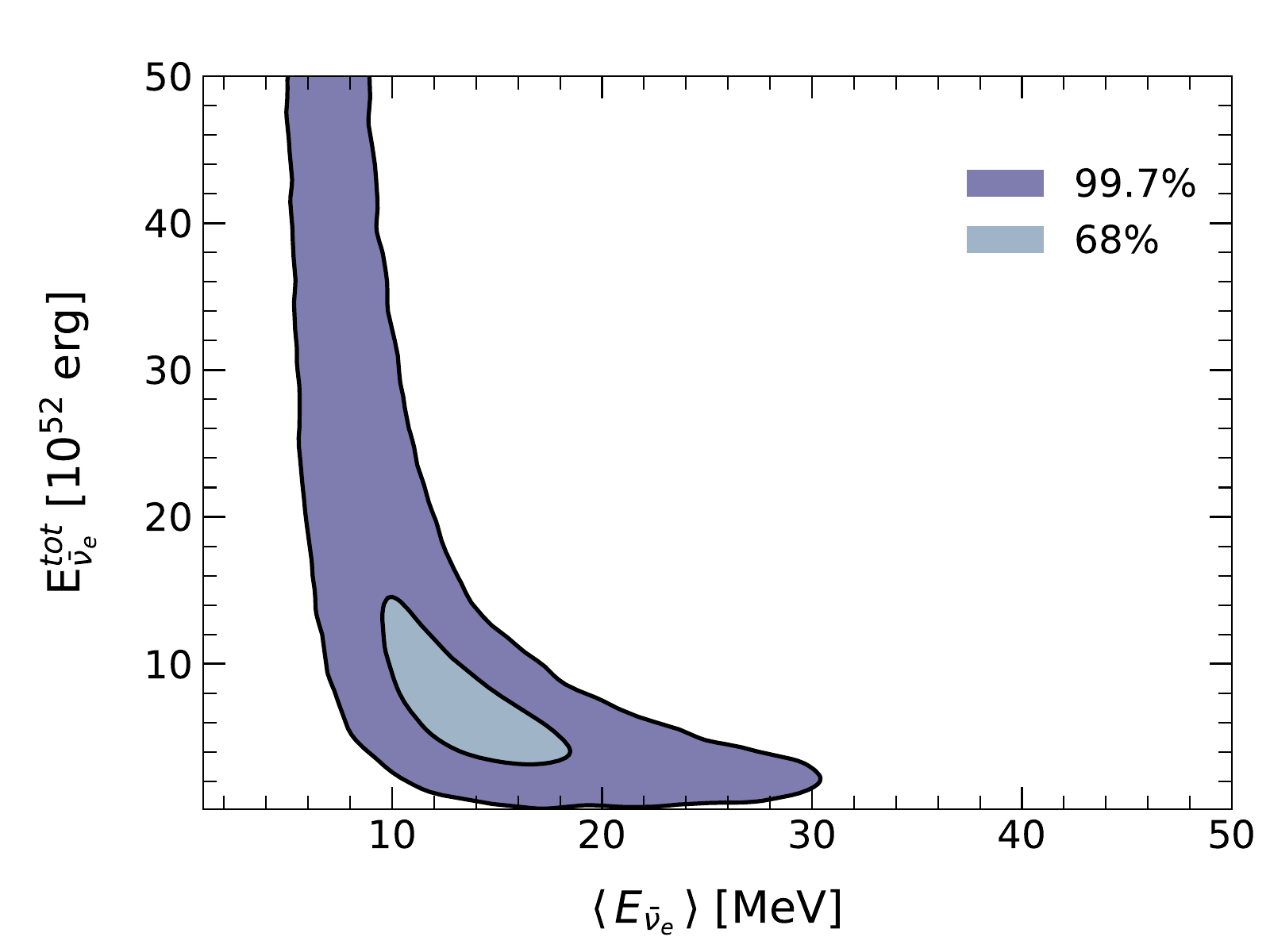}
    \caption{Posterior probability distribution functions of the supernova neutrino parameters obtained by observing the DSNB in DUNE and Hyper-Kamiokande, assuming no oscillations. Left and right panels refer to average and total energy of $\nu_e$ and $\bar{\nu}_e$, respectively.}
    \label{fig:only_DSNB}
\end{figure*}

\begin{figure*}
    \centering
    \text{\scriptsize \hspace{6mm} No oscillations, $C^{238}=10^{-11}$}\par
    \includegraphics[trim={0cm 0cm 0cm 6mm},clip,width=0.32\linewidth]{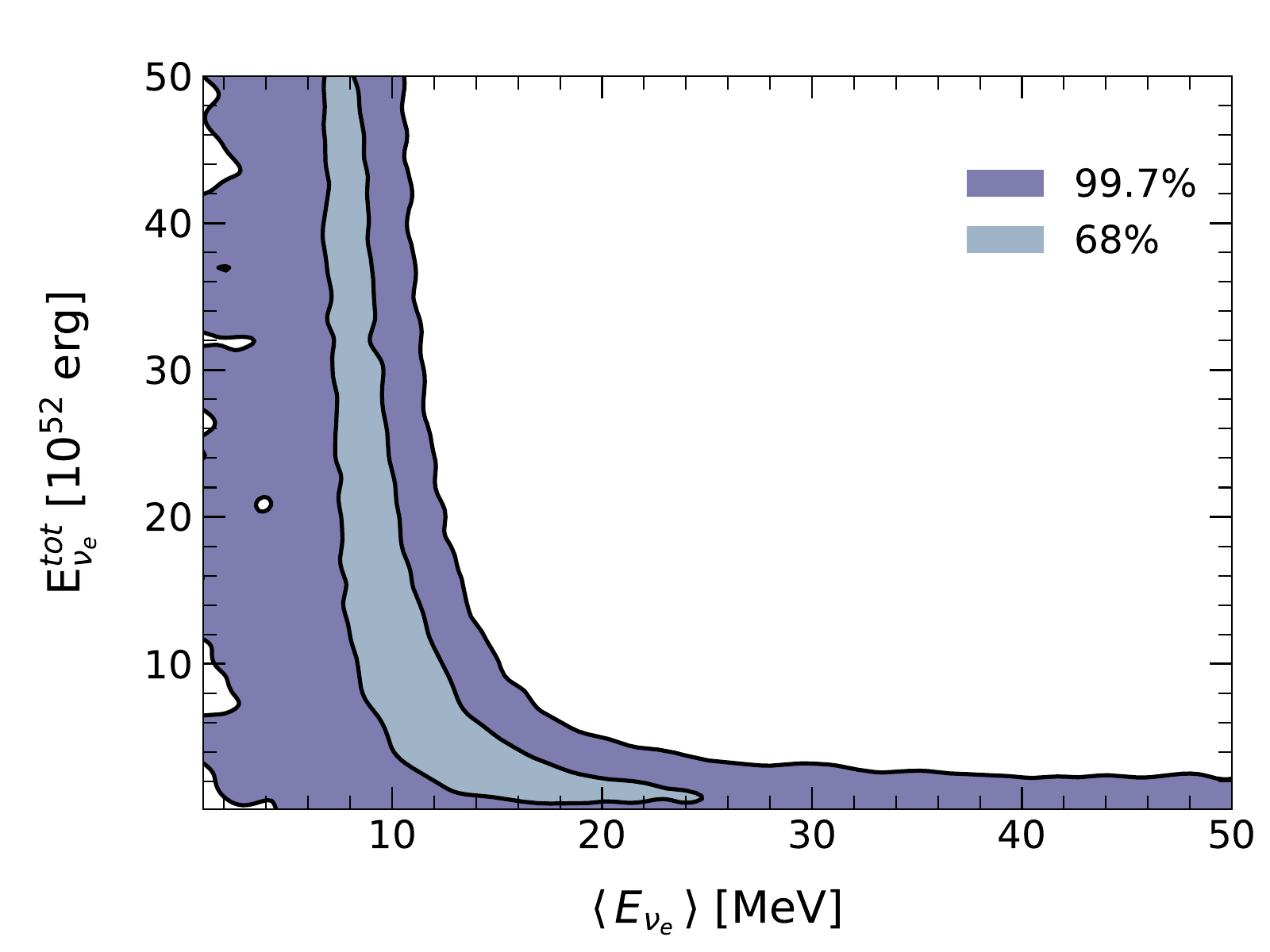}
    \includegraphics[trim={0cm 0cm 0cm 6mm},clip,width=0.32\textwidth]{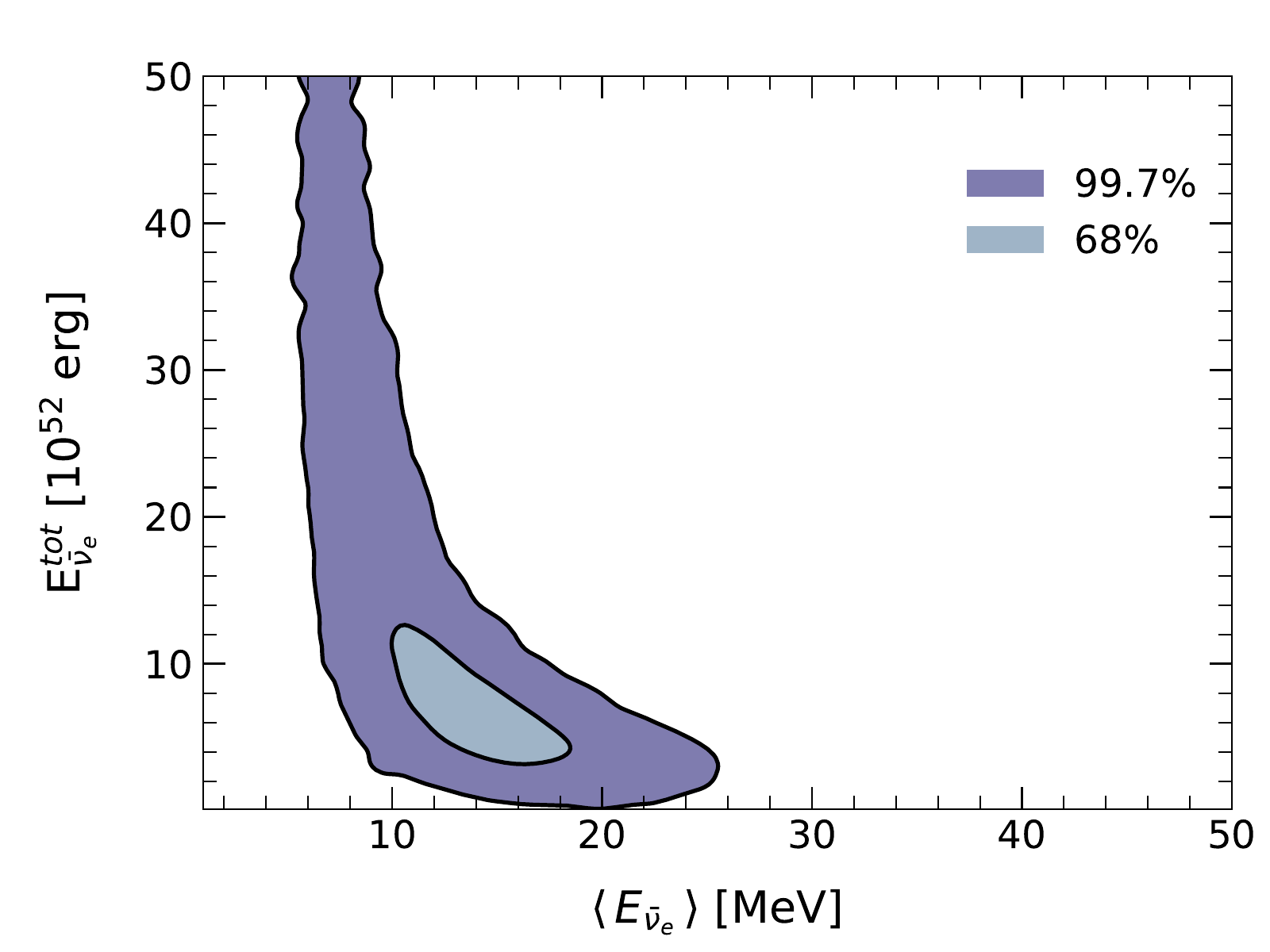}
    \includegraphics[trim={0cm 0cm 0cm 6mm},clip,width=0.32\textwidth]{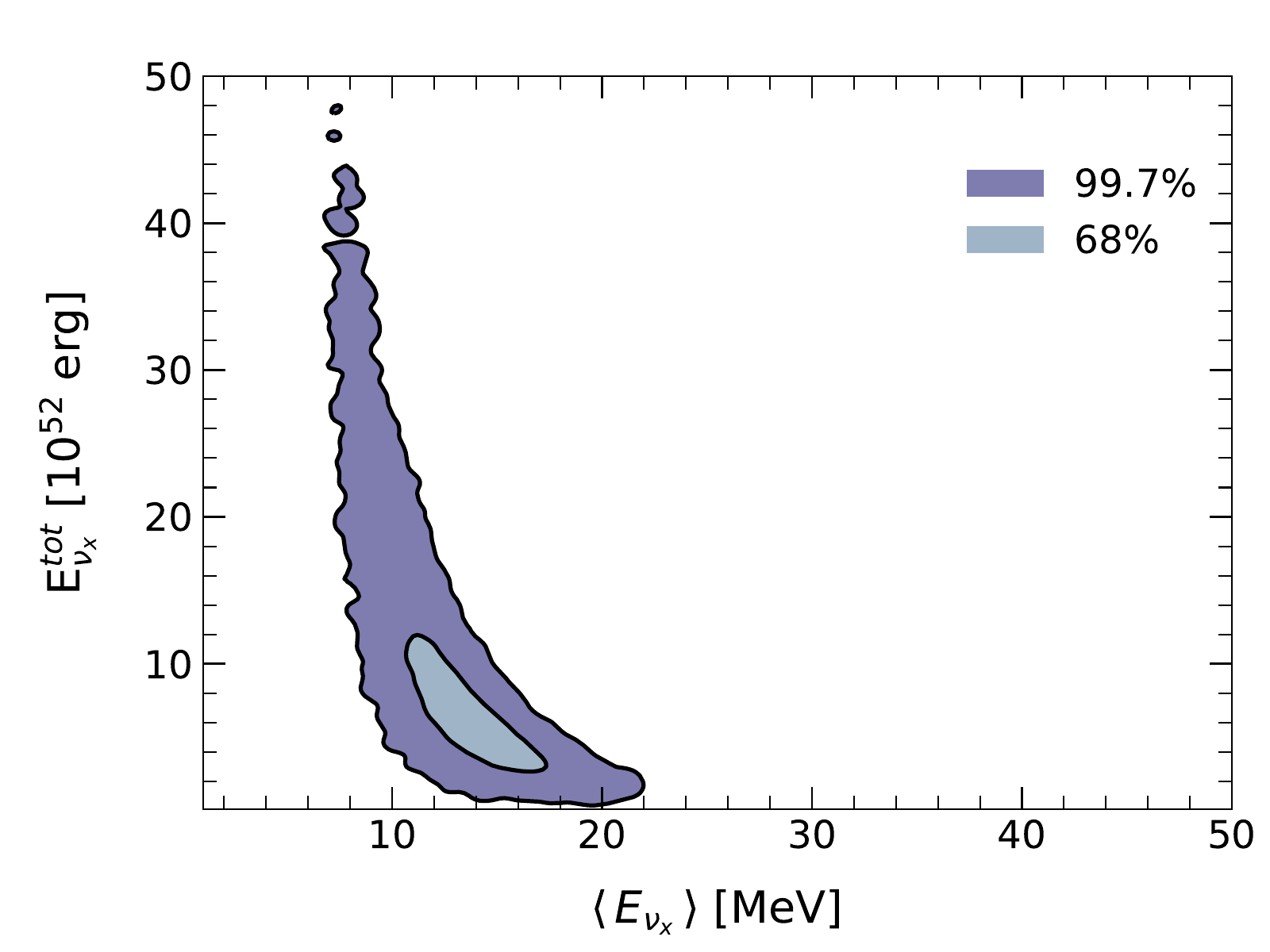}
    \caption{Posterior probability distribution functions of the supernova neutrino parameters obtained assuming no oscillations and $C^{238}=10^{-11}$, using both paleo detectors and DSNB observations in Hyper-Kamiokande and DUNE. Left, central and right panels refer to average and total energy of $\nu_e$, $\bar{\nu}_e$ and $\nu_x$, respectively.}
    \label{fig:standard_background_no_oscillations}
\end{figure*}

\begin{figure*}
    \centering
    \text{\scriptsize \hspace{-0.8cm} Normal ordering, $C^{238}=10^{-11}$ \hspace{5.0cm} Inverted ordering, $C^{238}=10^{-11}$}\par
    \includegraphics[trim={0cm 0cm 0cm 6mm},clip,width=0.48\linewidth]{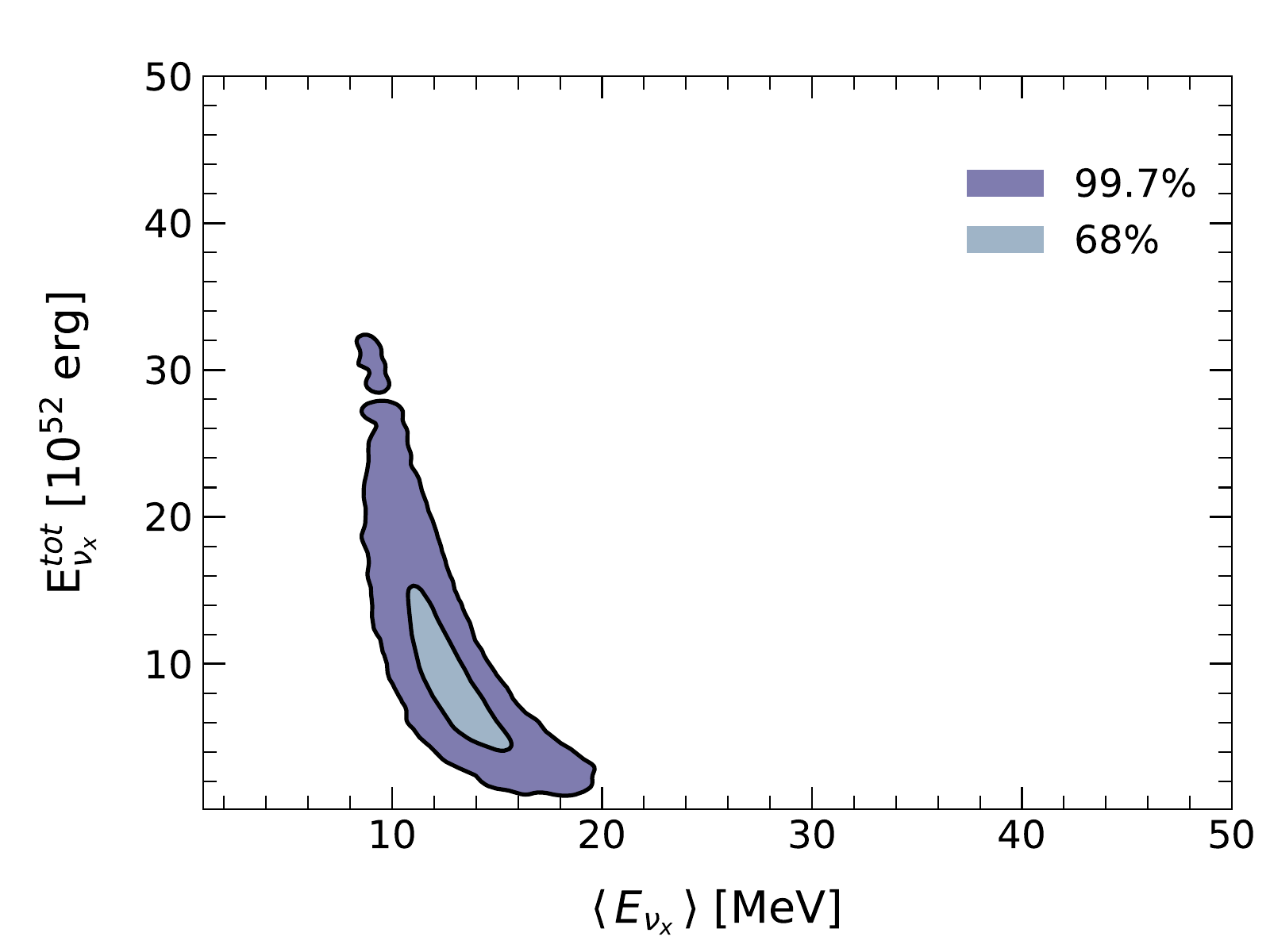}
    \includegraphics[trim={0cm 0cm 0cm 6mm},clip,width=0.48\textwidth]{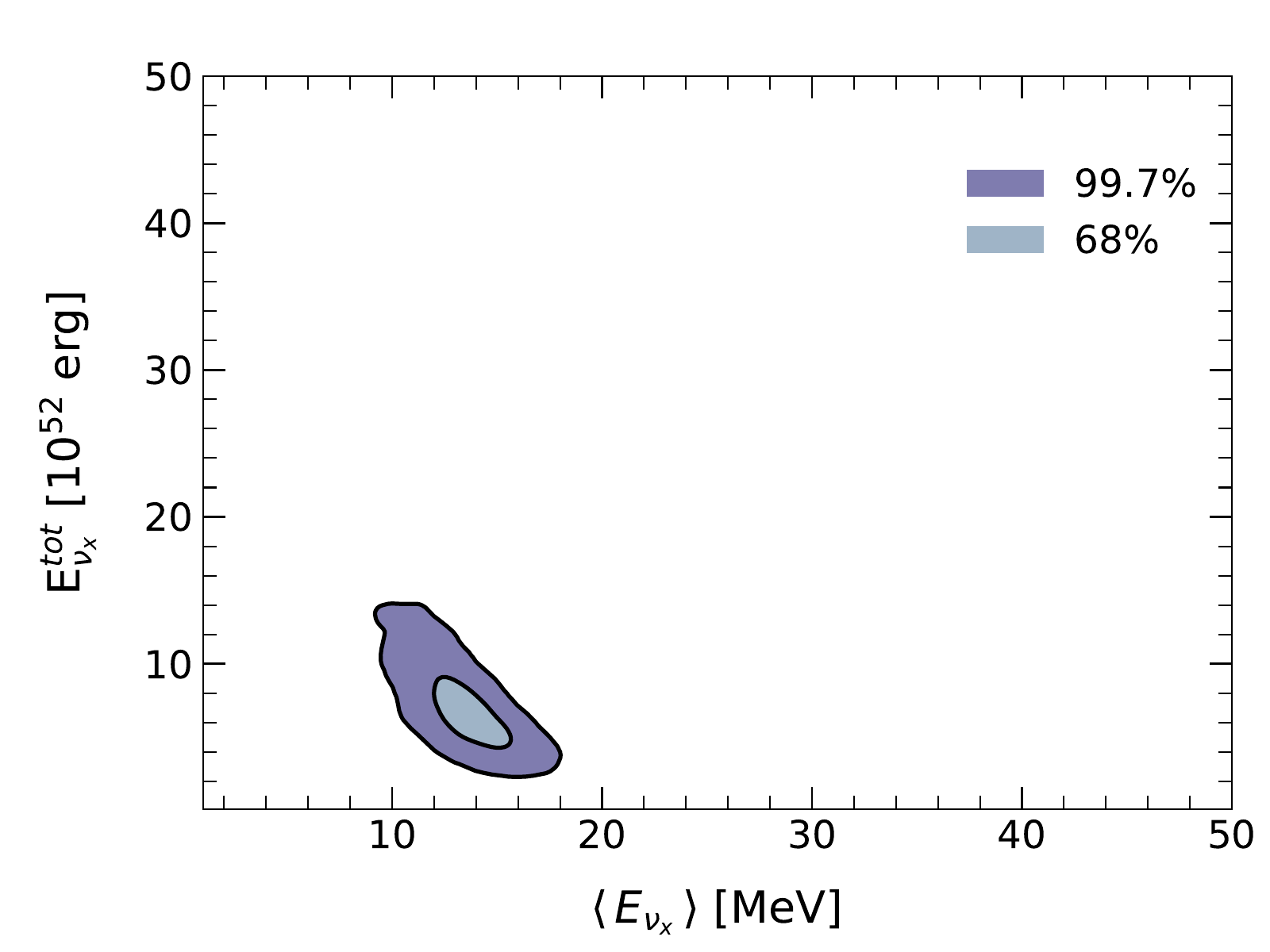}
    \caption{Same as figure~\ref{fig:standard_background_no_oscillations}, but for normal ordering (left panel) and inverted ordering (right panel) of $\nu_x$.}
    \label{fig:standard_background_oscillated}
\end{figure*}

Figure~\ref{fig:only_DSNB} displays the allowed regions at 68\% and 99.7\% confidence levels for the average energy and total energy of $\nu_e$ and $\bar{\nu}_e$ assuming no oscillations, obtained using only the measurement of DSNB neutrinos in DUNE and Hyper-Kamiokande for 20-years of operation. A degeneracy between average energy and total energy is observed, which is due to the fact that the statistics of the signal events is proportional to the product $\langle E_{\nu_\alpha}\rangle^2 E_{\nu_{\alpha}}^{\rm tot}$. A better precision is reached for Hyper-Kamiokande (right panel) due to its larger statistics. Figure~\ref{fig:standard_background_no_oscillations} displays the allowed regions at 68 and 99.7\% confidence levels for the average energy and total energy of $\nu_e$, $\bar{\nu}_e$ and $\nu_x$, assuming $C^{238}=10^{-11}$ and no oscillations, obtained using both paleo detectors and DSNB neutrinos in Hyper-Kamiokande and DUNE. The main result here is that the best precision is obtained for the $\nu_x$ parameters since the $\nu_x = \{ \nu_\mu, \bar{\nu}_\mu, \nu_\tau, \bar{\nu}_\tau\}$ contribute roughly four times more to the total neutrino flux than $\nu_e$ or $\bar{\nu}_e$, see eq.~\eqref{eq:flavor_sum}. In particular, the values $E^{\rm tot}_{\nu_x}=0$ and $\langle E_{\nu_x}\rangle=0$ are disfavored at more than 99.7\% confidence level. Moreover, the constraints on the parameters for the electron flavors are slightly improved with respect to the case with only DSNB neutrinos. Figure~\ref{fig:standard_background_oscillated} is equivalent to figure~\ref{fig:standard_background_no_oscillations}, but it includes oscillation effects for both normal and inverted mass ordering and, for brevity, we show only the results for the constraints on the $\nu_x$ parameters. For both mass orderings, the bounds on $\nu_x$ parameters are slightly improved with respect to the no oscillation case. Although oscillations do not affect the neutral current coherent scattering signal in paleo detectors, oscillations do lead to signals from the $\nu_x$ component in either Hyper-Kamiokande or DUNE, depending on the mass ordering under consideration. The best precision is reached for inverted ordering, since in this case, $\nu_x$ can be observed in Hyper-Kamiokande, which has a larger statistics compared to DUNE.

\begin{figure*}
    \centering
    \text{\scriptsize \hspace{0.7cm} No oscillations, $C^{238}=10^{-12}$ \hspace{0.9cm} Normal ordering, $C^{238}=10^{-12}$ \hspace{0.8cm} Inverted ordering, $C^{238}=10^{-12}$}\par
    \includegraphics[trim={0cm 0cm 0cm 6mm},clip,width=0.32\linewidth]{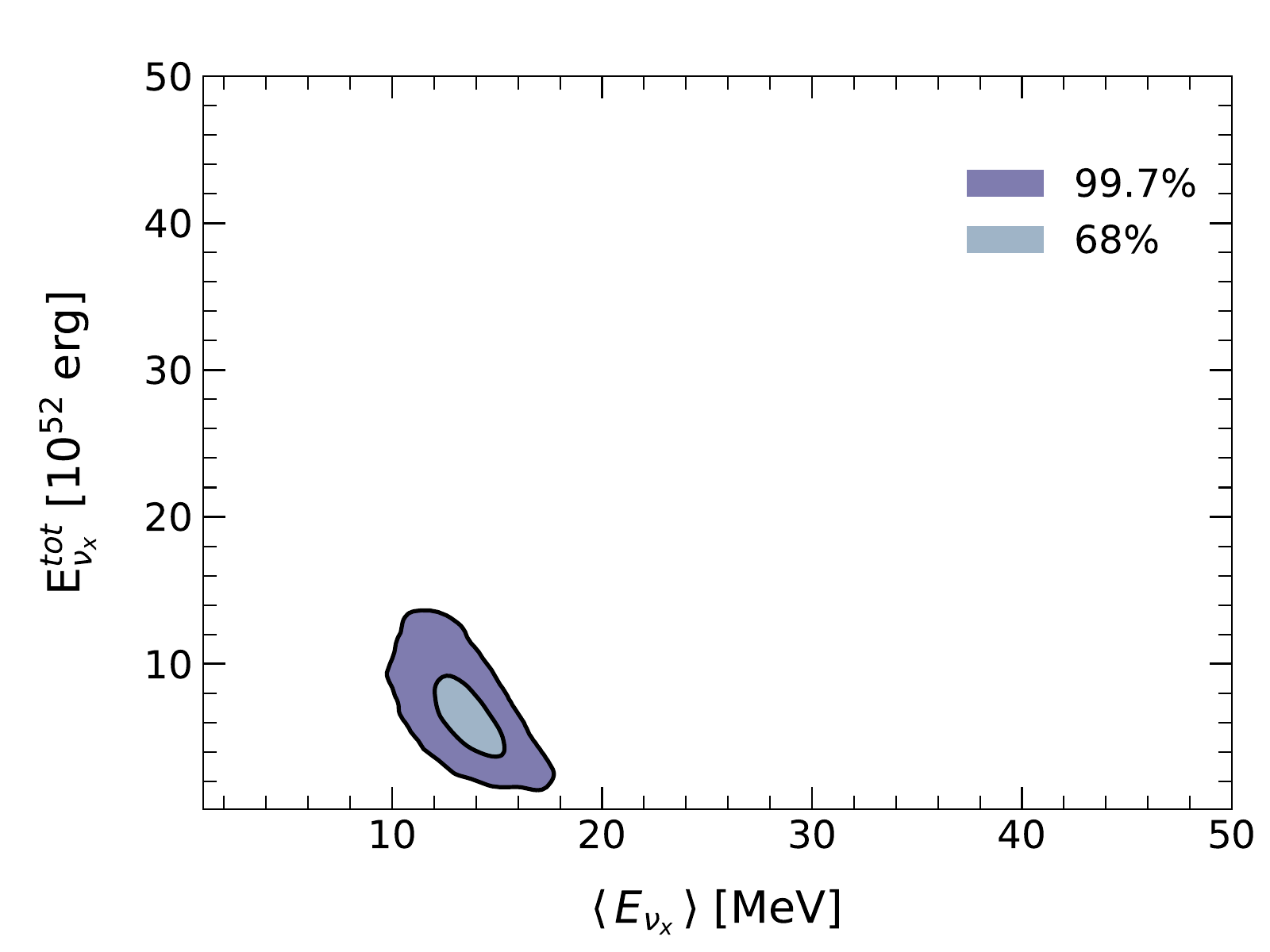}
    \includegraphics[trim={0cm 0cm 0cm 6mm},clip,width=0.32\linewidth]{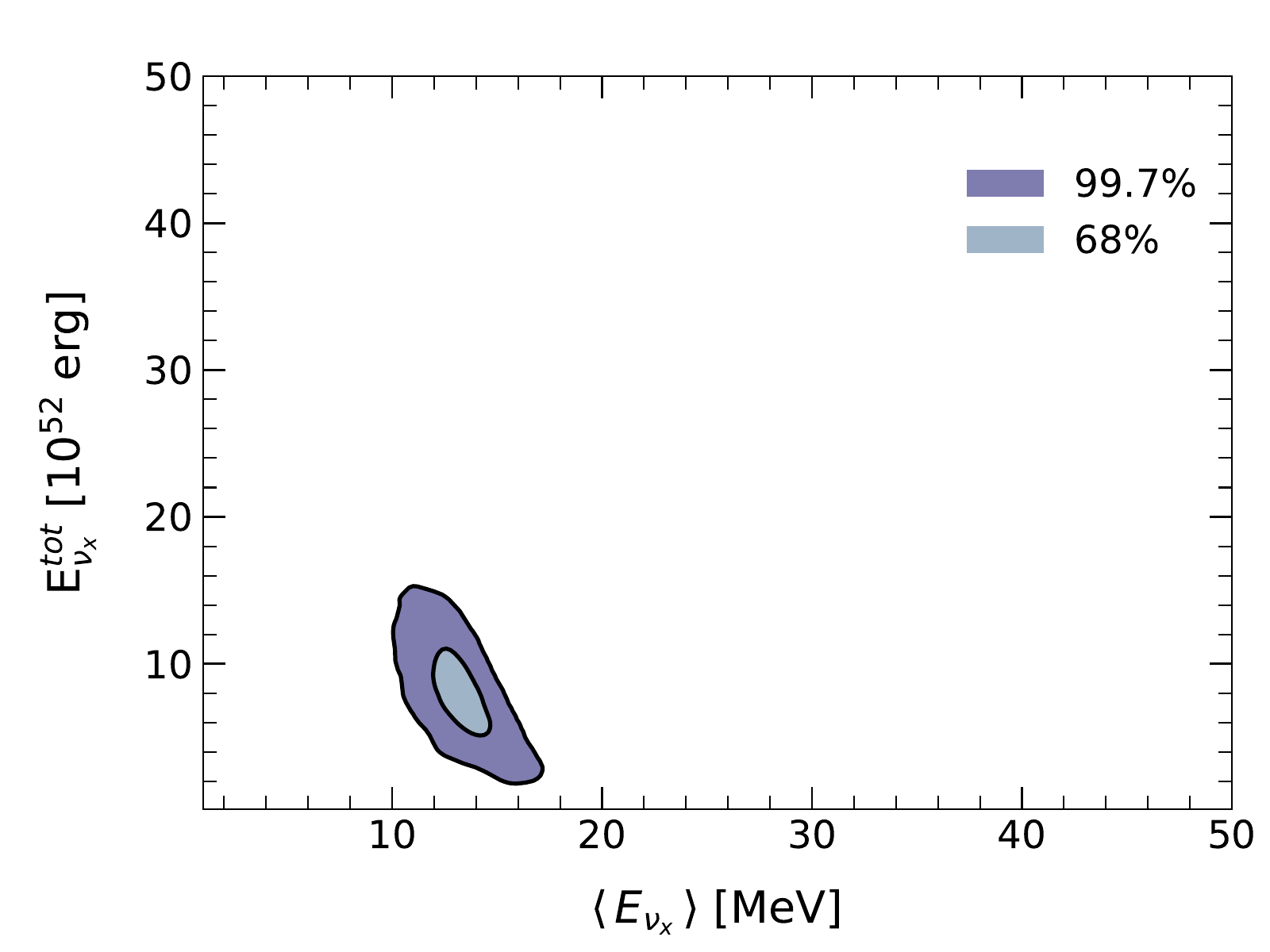}
    \includegraphics[trim={0cm 0cm 0cm 6mm},clip,width=0.32\linewidth]{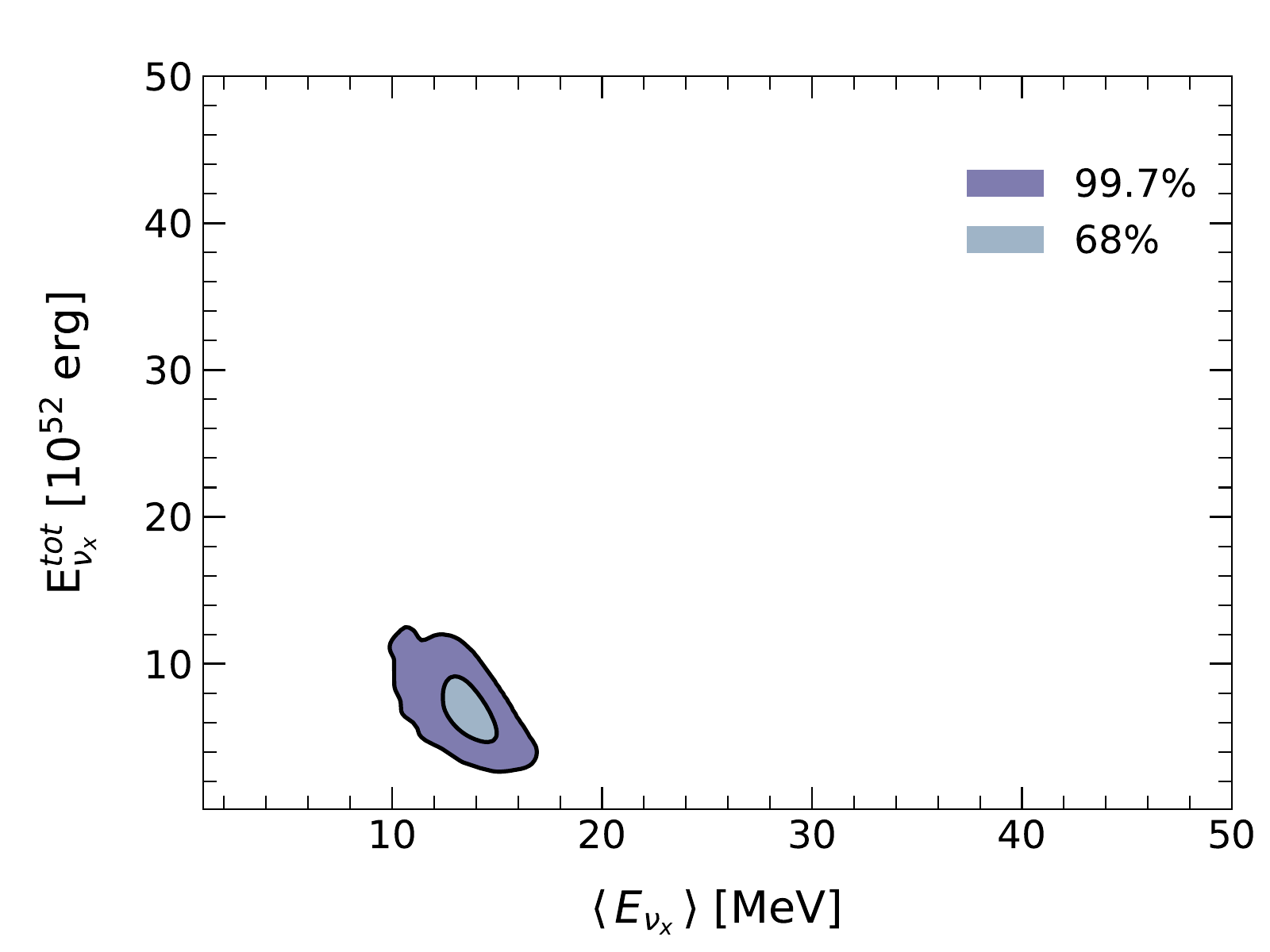}
    \caption{Posterior probability distribution functions of the supernova neutrino parameters obtained assuming no $C^{238}=10^{-12}$. Left, central and right panels refer to the case of no oscillations, normal ordering and inverted case of $\nu_x$, respectively.}
    \label{fig:reduced_background}
\end{figure*}

In figure~\ref{fig:reduced_background} we show the projected constraints on the $\nu_x$ parameters for the three different assumptions of the flavor oscillations considered above but for a more optimistic $^{238}$U concentration of $C^{238} = 10^{-12}\,$g/g in the paleo detectors rather than the $C^{238} = 10^{-11}\,$g/g assumed in figures~\ref{fig:standard_background_no_oscillations} and~\ref{fig:standard_background_oscillated}. Comparing these figures we can see that using paleo detectors with uranium concentration of $C^{238} = 10^{-12}\,$g/g would allow to constrain $\langle E_{\nu_i} \rangle$ and $E_{\nu_i}^{\rm tot}$ with (relative) precision $\lesssim 10\%$ rather than the few tens of percent precision we project for $C^{238} = 10^{-11}\,$g/g. This improvement is due to the reduction of the dominant background in paleo detectors, the (elastic) scattering of neutrons originating from the $^{238}$U decay chain.

If we modify the assumption of a 10\% prior on both the galactic supernova rate and the normalization of the DSNB, our results are significantly affected (not shown). In particular, when increasing the width of these priors to 20\%, the 99.7\% confidence level region extends towards $\langle E_{\nu_x}\rangle=0$ and large values of $E_{\rm tot}^{\nu_x}$. This happens because widening this prior loosens the link between the DSNB measurements in Hyper-Kamiokande and DUNE with the paleo-detector measurement of neutrinos from galactic core-collapse events, making it possible for the contribution of $\nu_x$ to be compensated by $\nu_e$ and $\bar{\nu}_e$. On the other hand, increasing the uncertainty on the normalization of backgrounds for DSNB experiments does not appreciably affect the final constraints on $\langle E_{\nu_x}\rangle$ and $E_{\nu_x}^{\rm tot}$.

A final remark is in order. Recently, the possibility of measuring the DSNB in conventional dark matter detectors and extracting the $\nu_x$ parameters has been carefully studied in Ref.~\cite{Suliga:2021hek}. Analogously to the signal in paleo detectors, in conventional dark matter detectors the supernova neutrino signal is dominated by background, but the signal-to-noise ratio is smaller for two reasons: Conventional dark matter detectors have much smaller exposures than paleo detectors (e.g, the DARWIN proposal has an envisaged exposure on the order of $\varepsilon \sim 100\,$t\,yr~\cite{DARWIN:2016hyl}). Second, conventional dark matter detectors would be sensitive to DSNB neutrinos, while paleo detectors could measure the (time-averaged) signal from galactic supernova neutrinos; the time-averaged galactic supernova flux is roughly two orders of magnitude larger than the DNSB flux. This difference in the signal-to-noise ratio is the reason why in Ref.~\cite{Suliga:2021hek}, only prospective upper limits on $\nu_x$ parameters are provided, whereas in our work, we show goodness-of--fit regions. Nevertheless, the two approaches to measure the $\nu_x$ component are complementary. Paleo detectors allow only to probe galactic supernovae, whereas dark matter detectors can measure the DSNB, which includes extragalactic supernova neutrinos. Furthermore, neutrinos propagating over cosmological distances, like those of the DSNB, allow studying different fundamental physics (see, for example, Refs.~\cite{Jeong:2018yts,DeGouvea:2020ang}).

\section{Conclusions}
\label{sec:Conclusions}

Supernova neutrinos carry invaluable information about the astrophysical processes occurring deep inside a stellar core collapse. The few tens of events observed from SN1987a have already demonstrated the potential of detecting a supernova neutrino burst, despite the small statistics collected. Next generation experiments, such as Hyper-Kamiokande, DUNE and JUNO, would detect $\mathcal{O}(10^5)$ $\bar{\nu}_e$, $\mathcal{O}(10^3)$ $\nu_e$ and $\mathcal{O}(10^2)$ $\nu_x$ events, respectively, from a galactic supernova, thus covering all neutrino flavors with differing precision. However, only a few supernovae are expected to occur per century in our galaxy on average, and the only remedy is patience. A possible way around this problem is probing the DSNB, i.e., the integrated flux of neutrinos produced by all the supernova explosions occurring throughout the Universe. Another benefit of the DSNB is that it allows one to be sensitive to the entire population of supernovae and not only to a specific progenitor, which might have peculiar characteristics. Hyper-Kamiokande, DUNE and JUNO have the capability to detect the DSNB, but with much smaller statistics compared to a single galactic supernova and with a significant background to deal with. For $\bar{\nu}_e$ events in Hyper-Kamiokande, such background can be mitigated by adding gadolinium in the water Cherenkov detector, which can provide a 90\% efficiency in tagging inverse beta decay events induced by neutrinos. For $\nu_x$, the detection prospects are less promising, even considering dark matter detectors where the background is intrinsically lower. For instance, the expected number of DSNB events in XENONnT is $\mathcal{O}(10^{-3})$ per year~\cite{Suliga:2021hek}, whereas the background is expected to add up to a few events per year~\cite{XENON:2020kmp}. In this case only upper limits can be set on the parameters describing the $\nu_x$ flux, i.e., the average energy and total energy, $\langle E_{\nu_x}\rangle$ and $E_{\nu_x}^{\rm tot}$ \cite{Suliga:2021hek}.

In this work we have shown that one could {\it measure} the $\nu_x$ parameters using paleo-detectors. Differently from conventional detectors which measure neutrino events in real time, paleo detectors integrate events over time scales as large as a billion years. Thus, paleo detectors could measure the integrated neutrino flux from core-collapse supernovae in our Galaxy via the nuclear recoils induced by (flavor-blind) CE$\nu$NS reactions of neutrinos with the atomic nuclei in paleo detectors. Leveraging the gigayear exposure times, with only 100\,g of a epsomite paleo detector, we expect about 3$\times 10^3$ $\nu_x$ events and a background of  $2\times 10^5$ events, assuming a uranium-238 concentration of $C^{238}=10^{-11}\,$g/g. Although the signal-to-background ratio is similar to what is expected for DSNB events in dark matter detectors, the statistics is orders of magnitude higher, leading to a much better signal-to-noise ratio -- the ratio between signal events and the square root of background is 6.6$\sigma$. Combining such a measurement of the galactic supernova neutrino flux with measurements of the DSNB via charged current interactions in Hyper-Kamiokande and DUNE would allow to provide closed contours of the 68\% and 99.7\% confidence level regions for the parameters controlling the supernova neutrino spectra of all flavors, $\nu_e$, $\overline{\nu}_e$, and $\nu_x$. With a more optimistic assumption on the concentration of uranium, $C^{238}=10^{-12}\,$g/g, we project that the parameters $\langle E_{\nu_x}\rangle$ and $E_{\nu_x}^{\rm tot}$ could be measured with a precision of $\lesssim 10\%$ on each parameters. However, the precision on some of these supernova neutrino parameters can be lost for uncertainties on the DSNB and galactic supnernova rate larger than the 10\% assumed here. Current uncertainties on the cosmic star formation rate at low redshifts are already at the level of $\sim 20\%$~\cite{Hopkins:2006bw,Beacom:2010kk,Madau:2014bja}, and while reaching $\sim 10\%$ uncertainties is ambitious, upcoming surveys with, for example, the James Webb Space Telescope or the Vera C. Rubin Observatory are expected to lead to considerable improvements in our knowledge of the cosmic and local star formation rate in the near future. We emphasize that the use of paleo detectors is complementary to the one of dark matter detectors, since they are sensitive to a different population of supernova progenitors and potentially to different effects of physics beyond the Standard Model.

\section*{Acknowledgements}
SB is supported in part by NSF Grant PHY-2014215, DOE HEP QuantISED award \#100495, and the Gordon and Betty Moore Foundation Grant GBMF7946. 
The work of FC at Virginia Tech is supported by the U.S. Department of Energy under the award number DE-SC0020250 and DE-SC0020262. The work of FC at IFIC is supported by GVA Grant No.CDEIGENT/2020/003.
The work of SH is supported by the US Department of Energy under the award number DE-SC0020262 and NSF Grant numbers AST-1908960 and PHY-1914409. This work was supported by World Premier International Research Center Initiative (WPI Initiative), MEXT, Japan. 

\bibliographystyle{bibi}
\bibliography{Bibliography}

\end{document}